\documentclass[useAMS,usenatbib]{mn2e}

\usepackage[usenames,dvips]{color}
\usepackage{graphicx}
\usepackage{defs}
\usepackage{cite}

\title[Massive Milky Way Satellites in Cold and Warm Dark Matter: Dependence on Cosmology]{Massive Milky Way Satellites in Cold and Warm Dark Matter: Dependence on Cosmology}

\author[E. Polisensky and M. Ricotti]{E. Polisensky$^{1}$\thanks{E-mail: Emil.Polisensky@nrl.navy.mil} and M. Ricotti$^{2,3}$\footnotemark[1]\thanks{E-mail: ricotti@astro.umd.edu}\\
$^{1}$Naval Research Laboratory, Washington, D.C. 20375, USA\\
$^{2}$Department of Astronomy, University of Maryland, College Park, Maryland 20745, USA\\
$^{3}$Sorbonne Universit\'{e}s, Institut Lagrange de Paris (ILP), 98 bis Bouldevard Arago 75014 Paris, France}
\begin{document}

\date{\today}

\pagerange{\pageref{firstpage}--\pageref{lastpage}} \pubyear{2013}

\maketitle

\label{firstpage}

\begin{abstract}

We investigate the claim that the largest subhaloes in high resolution
dissipationless cold dark matter (CDM) simulations of the Milky Way are 
dynamically inconsistent with observations of its most luminous satellites. 
We find that the inconsistency is largely attributable to the large values of 
$\sigma_8$ and $n_s$ adopted in the discrepant simulations, producing 
satellites that form too early and therefore are too dense. 
We find the tension 
between observations and simulations adopting parameters 
consistent with WMAP9 is greatly diminished, 
making the satellites a sensitive test of CDM. 
We find the Via~Lactea~II halo to be atypical for haloes in a WMAP3 cosmology, 
a discrepancy that we attribute to its earlier
formation epoch than the mean for its mass.
We also explore warm dark matter (WDM) cosmologies for 1--4~keV thermal relics. 
In 1~keV cosmologies subhaloes have circular velocities at kpc scales 
$\sim 60\%$ lower than their CDM counterparts, but are reduced by only 
$10\%$ in 4~keV cosmologies.
Since relic masses $<$ 2-3~keV are ruled out by constraints from the 
number of Milky Way satellites and Lyman-$\alpha$ forest, 
WDM has a minor effect in reducing the densities of massive satellites. 
Given the uncertainties on the mass and formation epoch of the Milky Way, 
the need for reducing the satellite densities with baryonic effects or 
WDM is alleviated.
\end{abstract}


\section{Introduction}\label{sec:1}

The satellite galaxies of the Milky Way (MW), being the closest
extragalactic objects and indeed within the virial radius of the Milky
Way's extended halo of dark matter, are uniquely suited for testing
theories of galaxy formation and evolution and the nature of dark
matter. The MW satellites known before the Sloane Digital Sky Survey
(SDSS) numbered too few to account for predictions from N-body
simulations in $\Lambda$CDM cosmologies that were otherwise successful
in descrbing the abundances of galaxies in clusters and the large
scale features of the matter distribution \citep{kly1999,moo1999}.
The discovery of a population of fainter satellites in the SDSS and
more sophisticated simulations that account for supernova feedback and
the heating of the intergalactic medium (IGM) during 
reionization have alleviated this
problem by predicting a strong suppression of galaxy formation in low
mass haloes \citep{bul2000, RGSa2002, RGS2002, RGS2008}.

Recent work focusing on the brightest MW satellites has highlighted
dynamical discrepancies with high-resolution CDM simulations.
\citet{boy2011,boy2012} compared the most luminous satellites to
subhaloes in the Aquarius simulation suite of six Milky Way-sized
haloes. Abundance matching models set a one-to-one correspondence
between luminosity and dynamical mass and place the brightest
satellites in the largest subhaloes. However, the observed stellar
velocities cannot be reconciled with the velocity profiles of the
largest dark matter subhaloes in simulation. The most massive
satellites, either at the present epoch, the epoch of reionization, or 
over the complete infall
history, are too dense to be dynamically consistent with the Milky Way
satellites. Observations of the stellar velocity dispersions in the
bright satellites are consistent with dark matter haloes with maximum
circular velocities $< 25$~km~s$^{-1}$ while the Aquarius Milky Ways
have about 10 subhaloes each with $v_{max} > 25$~km~s$^{-1}$ that are
also not Magellenic Cloud analogues. Several solutions to this
problem have been proposed. Galaxy formation may be stochaistic on
dwarf spheroidal scales and the bright satellites do not reside in the
largest subhaloes \citep{boy2011, KatzR2013}. This requires abandoning
the monotonic relation between galaxy luminosity and halo mass that is
well-established for brighter galaxies.

Interestingly, in models in which some of the ultra-faint dwarfs are
fossils of the first galaxies \citep{RicottiG2005, BovillR09}, show
some tension with observations only at the bright end of the satellite
luminosity function \citep{BovillRa2011, BovillRb2011}. Simulations
that produce a numerous population of ultra-faint dwarfs also produce
an overabundance of bright dwarf satellites especially in the outer
parts of the Milky Way. However, this tension is eased by the expected
stripping of the extended primordial stellar population around bright
satellites.

The number of satellites of all size are known to be proportional to
the mass of the host halo \citep{kly1999}. \citet{wan2012} argue the
low velocities of the MW satellites may be an indication the MW is
less massive than typically thought. They show there is only a $5\%$
probability for a galaxy of mass $2\times10^{12} M_{\sun}$ to have 3
satellites or less with maximum circular velocities $> 30$~km~s$^{-1}$
but $40\%$ for a galaxy of mass $10^{12} M_{\sun}$. A low mass for the
Milky Way of $8\times10^{11} M_{\sun}$ is also favored in the work of
\citet{ver2013}. Direct measures of the MW mass typically focus on
stellar tracers of the inner halo or radial velocity measurements of
the MW satellites and give a range of virial mass $0.8-2.5 \times
10^{12} M_{\sun}$, we refer the reader to the references in
\citet{boy2012b} where observations of the spatial motion of Leo I are
used to constrain the mass of the Milky Way to $> 10^{12} M_{\sun}$ at
$95\%$ confidence.

\citet{saw2012} show the simulations can be reconciled with the
observations by including baryonic physics in the
simulations. Inclusion of baryonic physics removes gas from haloes
through supernova expulsion of the interstellar medium, prevention of
gas accretion through reionization heating of the IGM, and ram pressure 
stripping from satellites. Removal of baryons
from the dark matter haloes also reduces the potential well resulting
in less accretion of both gas and dark matter. They show dark matter
only simulations overpredict the subhalo abundance by $30\%$ at a mass
scale of $10^{10} M_{\sun}$ with an increasing number of subhaloes
with no gas or stars below this scale.

The influence of baryons was also studied by \citet{dic2011}. They
found that while satellites with low baryon fractions have lower
concentrations than their dark matter only counterparts, satellites
with high baryon fractions have higher central densities due to
adiabatic contraction. Satellites with high baryon fractions also tend
to have the largest maximum circular velocities. However, their recent
work \citep{dic2013} finds the subhalo density profiles are better
described by Einasto profiles than Navarro, Frenk, and White (NFW)
profiles \citep{nav1997} and that this reconciles the observations
with simulated satellites of similar luminosities. \citet{ver2013}
also find agreement with Einasto profiles. However, while the initial
work of \citet{boy2011} assumes NFW profiles their later work
\citep{boy2012} uses the subhalo circular velocity profiles directly
with no assumed form.

Another possibility is a change in the nature of the dark matter from
standard CDM assumptions of collisionless particles with low
intrinisic thermal velocities. \citet*{vog2012} simulated one of the
Aquarius Milky Way haloes in self-interacting dark matter models. The
ability of the dark matter particles to self-scatter leads to the
formation of subhaloes with constant density cores. The lower density
decreases the inner circular velocity profiles bringing the
simulations into agreement with the observations.

A truncation in the dark matter power spectrum was investigated as a
solution to the paucity of satellites by reducing the abundance of
haloes at subgalactic scales. One method for producing a truncated 
power spectrum is 
if the dark matter particles decoupled with relativistic velocities
early in the radiation dominated era and thereby able to stream out of
overdense regions before becoming nonrelativistic at a time before the
horizon had reached Galactic scales. The scale of the power spectrum
truncation in `warm' dark matter (WDM) is related to the mass of the
dark matter particle with lighter particles decoupling earlier and
able to stream longer.

Dwarf-scale haloes in WDM cosmologies form later and have lower
concentrations than haloes in CDM, offering a potential solution to
the dynamical discrepancies. \citet{lov2012} simulated one of the
Aquarius haloes in a 1 keV thermal relic WDM cosmology and 
showed the subhaloes have 
central densities and velocity profiles in agreement with the bright
MW satellites. 
In \citet{lov2013a} their work was extended to particle masses 1.4-2.3~keV.
Recently, one Milky Way-like halo was simulated in WDM at 2, 3, 
and 4~keV \citep{sch2013}. 
In this work we investigate the subhalo dynamics in 
four Milky Way-sized haloes in $1, 2, 3,$ and $4$~keV
cosmologies.

Another area potentially affecting the 
subhalo densities are the adopted cosmological parameters.
The Via Lactea~II (VL2) simulation \citep{die2007,die2008}, which 
adopted parameters
from the 3rd year release of the \textit{Wilkinson Microwave 
Anisotropy Probe} (WMAP), 
was found to give
similar results as the six Aquarius haloes adopting WMAP1 parameters.
However, reason to suspect the adopted cosmology is important comes from 
\citet{mac2008} who explored the dependence of halo concentration 
on the adopted cosmological model for field galaxies.
They fit NFW density profiles to the haloes in their simulations:
\begin{equation}\label{eqNFW}
\rho(r) = \frac{\delta_c \rho_{crit}}{(r/r_s)(1+r/r_s)^2},\\
\end{equation}
and determined the concentrations, $c_{200}=R_{200}/r_s$, where $R_{200}$ 
is the radius 
enclosing a density 200 times the critical density, $\rho_{crit}$.
They found the average concentration of dwarf-scale field haloes varies by
a factor of 1.55 between WMAP1 and WMAP3.
In this work we also examine the dependence of our CDM subhalo populations 
on the
adopted cosmological parameters. 

\section{Simulations}

All our simulations were conducted with the $N$-body cosmological
simulation code {\sevensize GADGET-2} \citep{spr2005} with
gravitational physics only and initial conditions generated with the
{\sevensize GRAFIC2} software package \citep{ber2001}.  We use the
high resolution simulations presented in \citet{pol2011} where two
Milky Way-sized haloes were simulated in a cubic box with comoving
side length of 90 Mpc, mass resolution of $9.2\times10^4 M_{\sun}$,
and a $275$~pc gravitational softening length. 
We refer to these haloes as the \textit{set A}
and \textit{set B} simulations. We also ran a high resolution
simulation of halo \textit{C8} from \citet{pol2011} 
with a $138$~pc softening length 
and refer to this as our \textit{set C} simulations.
Finally, we ran an additional 
\textit{set D} simulation of another Milky
Way-sized halo in a 67 Mpc comoving box with a mass resolution
$8.2\times10^4 M_{\sun}$ and gravitational softening length $196$~pc.

Table~\ref{tabCOSMO} lists sets of cosmological parameters from
measurements of the cosmic microwave background by WMAP
and the \textit{Planck} mission \citep{wmap1,spe2007,wmap5,lar2010,
jar2010,kom2011,wmap9,planck1}. 
``Bolshoi'' are the parameters from the Bolshoi simulation 
\citep{kly2010} which were chosen to be within $1\sigma$ of WMAP5, WMAP7, 
and consistent with the results of supernovae, and X-ray cluster surveys.
These parameters are within $1\sigma$ of WMAP9 except the value 
of $n_s$ which is 
within $1.7\sigma$. 
They are also within $1.2\sigma$ of Planck1 with the exceptions 
of $\Omega_m$ and $\Omega_\Lambda$ which 
are $2.2\sigma$ below Planck1.
The WMAP1 parameters are $2.4-4.1\sigma$ away from
Planck1 while $\sigma_8$ and $n_s$ are $3.4\sigma$ and 
$2.2\sigma$ above WMAP9, respectively.
In contrast, the value of $\sigma_8$ in WMAP3 is $3.5\sigma$ below 
WMAP9 and Planck1.

\begin{figure}
\includegraphics*[scale=0.36,angle=270]{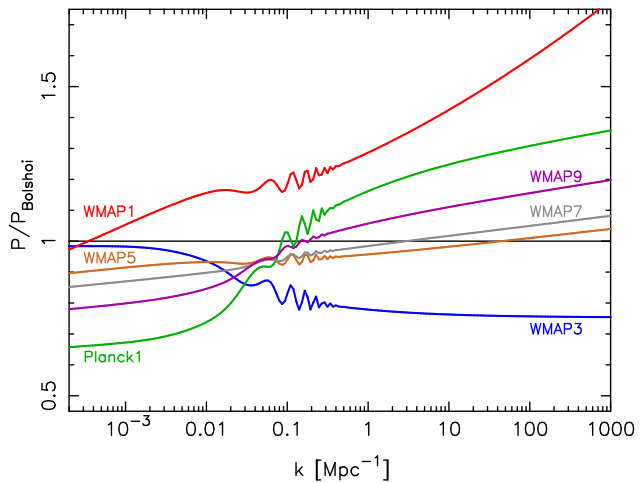}
\caption{Power spectra for CDM cosmologies normalized by the 
Bolshoi power spectrum.\label{figPS1}}
\end{figure}

Figure~\ref{figPS1} shows the linear power spectra for the parameters listed in 
Table~\ref{tabCOSMO} normalized by the Bolshoi power spectrum. 
On the scale of the dwarfs ($k\sim10$ Mpc$^{-1}$) the power varies greatly across
cosmologies with WMAP1 and WMAP3 representing the extremes of high and 
low power.
The Bolshoi parameters, however, represent a conservative estimate of 
the power on dwarf scales 
while being consistent with the latest CMB measurements from WMAP 
and \textit{Planck}.

To investigate the dependence of satellite densities on cosmology we
ran CDM simulations for each of our four sets adopting WMAP1, WMAP3, 
and Bolshoi parameters
with the CDM transfer function from \citet{eis1997}.
The box size and softening lengths were scaled in each simulation to 
keep the mass resolution constant.
A series of low resolution tests of the \textit{set B} halo were also 
run, these are described
in the next section.

For our investigation of warm dark matter we used the warm dark matter 
transfer function 
given by \citet*{bod2001} valid for particles in thermal equilibrium 
at the time 
of their decoupling, such as the gravitino. We adopted Bolshoi 
parameters and ran 
simulations for particle masses of 1, 2, 3, and 4~keV for each halo.

\begin{table}
\caption{Cosmological parameters.\label{tabCOSMO}}
\begin{center}
\begin{tabular}{l c c c c c c}    
\hline
 Name & $\Omega_{m}$ & $\Omega_{\Lambda}$ & $\Omega_{b}$ & $h$ & $\sigma_{8}$ & $n_{s}$ \\
\hline
\hline
 WMAP1 & 0.25 & 0.75 & 0.045 & 0.73 & 0.90 & 1.0\\
 WMAP3 & 0.238 & 0.762 & 0.040 & 0.73 & 0.74 & 0.951\\
 WMAP5 & 0.258 & 0.742 & 0.0441 & 0.72 & 0.796 & 0.963\\
 WMAP7 & 0.267 & 0.733 & 0.0449 & 0.71 & 0.801 & 0.963\\
 WMAP9 & 0.282 & 0.718 & 0.0461 & 0.70 & 0.817 & 0.964\\
 Planck1 & 0.317 & 0.683 & 0.0486 & 0.67 & 0.834 & 0.962\\
 Bolshoi & 0.27 & 0.73 & 0.0469 & 0.70 & 0.82 & 0.95\\
\hline
\end{tabular}
\end{center}
\end{table}

We used version 1.0 of the {\sevensize AMIGA's Halo Finder (AHF)}
software \citep{kno2009} to identify the Milky Way haloes and their
gravitationally bound subhaloes after iteratively removing unbound
particles. Table~\ref{tabMW} summarizes the properties calculated by
AHF for our simulated Milky Ways at $z=0$. We write $R_{100}$ to mean
the radius enclosing an overdensity $100$ times $\rho_{crit}$. 
The mass and number of particles inside $R_{100}$ are $M_{100}$
and $N_{100}$, respectively; $v_{max} =\mbox{max(}v_{circ}\mbox{)}$ is
the maximum circular velocity of the halo occuring at a radius
$R_{max}$, and $v_{circ}^2=GM(<r)/r$. 
Also given is the NFW $c_{200}$ concentration for each halo determined from:
\begin{equation}\label{eqC200}
\left(\frac {v_{max}}{v_{200}}\right)^2 = 0.2162\mbox{ } c_{200}/f(c_{200}),\\
\end{equation}
where $f(c) = \ln(1+c) - c/(1+c)$.

We also ran the {\sevensize SUBFIND} program \citep{spr2001} on 
the \textit{set B} WMAP3 data and
found excellent agreement with the results from {\sevensize AHF}.

\begin{figure}
\includegraphics*[scale=0.34,angle=270]{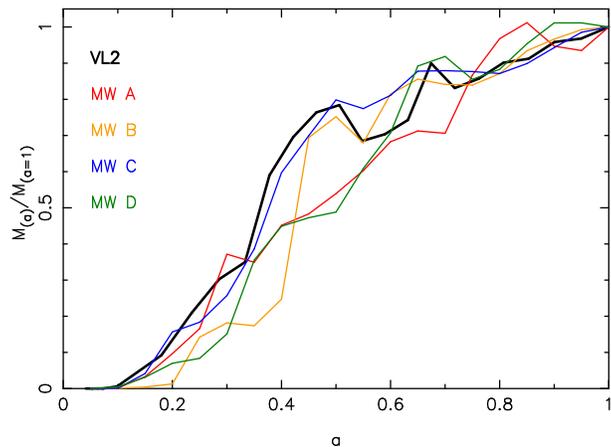}
\caption{Mass growth histories of simulated Milky Way haloes as a 
  function of scale factor, $a$.\label{figMH}}
\end{figure}

We saved snapshots of the particle information every 0.05 change 
in the universal scale 
factor, $a = (1+z)^{-1}$, for simulations adopting Bolshoi and WMAP1 parameters. 
Figure~\ref{figMH} shows the mass growth of each of our MW haloes and 
the VL2 halo 
as a function of $a$.
The masses are normalized to the halo mass at $a=1$.
The {\sevensize MergerTree} tool 
in {\sevensize AHF} was used to construct merger trees for all 
identified haloes.
This allows determination of $v_{infall}$ for each subhalo, the maximum 
value of $v_{max}$ 
over a halo's formation and accretion history: 
$v_{infall} = \mbox{max(}v_{max}(z)\mbox{)}$.
We follow the work of \citet{boy2011} and consider subhaloes
within 300 kpc of our Milky Way centres. 
We similarly identify subhaloes with $v_{max} > 40$ km s$^{-1}$ 
and $v_{infall} > 60$ km s$^{-1}$ as hosts of Magellanic Cloud analogues.

\begin{table*}
\caption{Properties of simulations and Milky Way haloes at $z=0$.\label{tabMW}}
\begin{center}
\begin{tabular}{l c c c c c c c r}    \hline
 \multicolumn{2}{c}{Cosmology} & $m_{res}$ & $M_{100}$ & $R_{100}$ & $v_{max}$ & $R_{max}$ & $N_{100}$ & $c_{200}$ \\
 & & [$M_{\sun}$] & [$10^{12}M_{\sun}$] & [kpc] & [km s$^{-1}$] & [kpc] & &\\
\hline
\hline
\multicolumn{9}{c}{\textit{Set A}}\\
\hline
\hline
 CDM & WMAP1 & $9.17\times10^4$ & 2.1119 & 324.233 & 214.78 & 39.849 & $23,028,026$ & 9.68\\
 CDM & Bolshoi & $9.17\times10^4$ & 1.9803 & 326.357 & 198.97 & 55.243 & $21,560,499$ & 8.38\\
 CDM & WMAP3 & $9.17\times10^4$ & 1.8410 & 309.740 & 192.28 & 41.027 & $20,074,556$ & 7.77\\
 4 keV & Bolshoi & $9.17\times10^4$ & 1.9644 & 325.486 & 198.11 & 50.414 & $21,387,017$ & 8.25\\
 3 keV & Bolshoi & $9.17\times10^4$ & 1.9724 & 325.929 & 197.05 & 54.900 & $21,474,003$ & 7.99\\
 2 keV & Bolshoi & $9.17\times10^4$ & 2.0061 & 327.771 & 197.54 & 39.871 & $21,874,542$ & 7.96\\
 1 keV & Bolshoi & $9.17\times10^4$ & 2.0197 & 328.514 & 199.24 & 58.943 & $22,022,816$ & 8.04\\
\hline
\hline
\multicolumn{9}{c}{\textit{Set B}}\\
\hline
\hline
 CDM & WMAP1 & $9.17\times10^4$ & 2.0873 & 322.973 & 210.02 & 67.068 & $22,760,127$ & 9.01\\
 CDM & Bolshoi & $9.17\times10^4$ & 1.9271 & 323.414 & 194.90 & 82.086 & $21,012,806$ & 7.81\\
 CDM & WMAP3 & $9.17\times10^4$ & 1.7540 & 304.781 & 194.62 & 79.767 & $19,125,479$ & 8.29\\
 4 keV & Bolshoi & $9.17\times10^4$ & 1.9193 & 322.971 & 194.19 & 74.900 & $20,928,496$ & 7.69\\
 3 keV & Bolshoi & $9.17\times10^4$ & 1.9224 & 323.157 & 193.65 & 77.500 & $20,962,535$ & 7.64\\
 2 keV & Bolshoi & $9.17\times10^4$ & 1.9242 & 323.257 & 194.53 & 79.500 & $20,981,724$ & 7.90\\
 1 keV & Bolshoi & $9.17\times10^4$ & 1.8804 & 320.771 & 195.23 & 84.286 & $20,503,730$ & 8.06\\
\hline
\hline
\multicolumn{9}{c}{\textit{Set C}}\\
\hline
\hline
 CDM & WMAP1 & $9.17\times10^4$ & 2.4195 & 339.274 & 231.42 & 44.932 & $26,240,319$ & 11.13\\
 CDM & Bolshoi & $9.17\times10^4$ & 2.3259 & 344.343 & 215.81 & 58.943 & $25,211,233$ & 9.05\\
 CDM & WMAP3 & $9.17\times10^4$ & 1.9887 & 317.808 & 203.42 & 56.164 & $21,645,271$ & 8.72\\
 4 keV & Bolshoi & $9.17\times10^4$ & 2.3195 & 344.029 & 215.03 & 56.900 & $25,152,203$ & 8.95\\
 3 keV & Bolshoi & $9.17\times10^4$ & 2.3194 & 344.014 & 215.25 & 57.100 & $25,153,016$ & 9.01\\
 2 keV & Bolshoi & $9.17\times10^4$ & 2.3113 & 343.614 & 214.40 & 61.529 & $25,070,237$ & 8.88\\
 1 keV & Bolshoi & $9.17\times10^4$ & 2.2607 & 341.086 & 210.94 & 64.857 & $24,563,114$ & 8.61\\
\hline
\hline
\multicolumn{9}{c}{\textit{Set D}}\\
\hline
\hline
 CDM & WMAP1 & $8.21\times10^4$ & 1.8164 & 308.342 & 190.95 & 67.027 & $22,135,114$ & 7.29\\
 CDM & Bolshoi & $8.21\times10^4$ & 1.5944 & 303.614 & 176.26 & 69.057 & $19,429,510$ & 6.80\\
 CDM & WMAP3 & $8.21\times10^4$ & 1.2575 & 272.781 & 164.27 & 50.164 & $15,323,846$ & 6.56\\
 4 keV & Bolshoi & $8.21\times10^4$ & 1.5930 & 303.526 & 176.62 & 75.414 & $19,412,993$ & 6.77\\
 3 keV & Bolshoi & $8.21\times10^4$ & 1.5875 & 303.171 & 176.38 & 74.143 & $19,345,715$ & 6.78\\
 2 keV & Bolshoi & $8.21\times10^4$ & 1.5548 & 301.086 & 175.73 & 75.729 & $18,947,343$ & 6.81\\
 1 keV & Bolshoi & $8.21\times10^4$ & 1.4998 & 297.486 & 171.97 & 79.514 & $18,276,956$ & 6.39\\
\hline
\hline
\multicolumn{9}{c}{\textit{Set B Low Resolution Tests}}\\
\hline
\hline
 CDM & WMAP1 & $7.34\times10^5$ & 2.3249 & 334.795 & 221.67 & 68.795 & $3,168,819$ & 9.56\\
 CDM sm & WMAP1 & $5.92\times10^5$ & 1.8899 & 312.452 & 208.53 & 73.630 & $3,192,628$ & 9.92\\
 CDM sm hi $z_i$ & WMAP1 & $5.92\times10^5$ & 1.9162 & 313.890 & 212.12 & 66.233 & $3,237,093$ & 10.63\\
 CDM & Planck1 & $7.34\times10^5$ & 2.3463 & 355.582 & 215.70 & 71.209 & $3,198,000$ & 10.02\\
 CDM & WMAP9 & $7.34\times10^5$ & 2.1919 & 337.600 & 210.15 & 78.429 & $2,987,609$ & 9.29\\
 CDM & Bolshoi & $7.34\times10^5$ & 2.0793 & 331.714 & 205.26 & 77.943 & $2,834,081$ & 8.91\\
 CDM & WMAP3 & $7.34\times10^5$ & 1.7650 & 305.411 & 191.01 & 98.288 & $2,405,721$ & 7.45\\
 CDM hi $z_i$ & WMAP3 & $7.34\times10^5$ & 1.9375 & 315.055 & 198.24 & 82.740 & $2,640,759$ & 7.92\\
\hline
\end{tabular}
\end{center}
\end{table*}

We compare our simulated subhaloes to the MW dwarf spheroidal
satellites with luminosities $L_V > 10^5 L_{\sun}$. \citet{walk2009}
and \citet{wolf2010} show line-of-sight velocity measurements provide
good constaints on the dynamical masses of dispersion-supported
galaxies like the MW dwarfs spheroidals. The Magellanic Clouds are
excluded from our observation sample as they are irregular type
galaxies. The Sagitarius dwarf is also excluded because it is
undergoing disruption and far from equilibrium. Our observed sample
consists of nine galaxies: Canes Venatici I, Carina, Draco, Fornax,
Leo I, Leo II, Sculptor, Sextans, and Ursa Minor.

\section{Results}

\subsection{Cold Dark Matter}\label{sec:3}

\begin{figure*}
\includegraphics*[scale=0.62,angle=270]{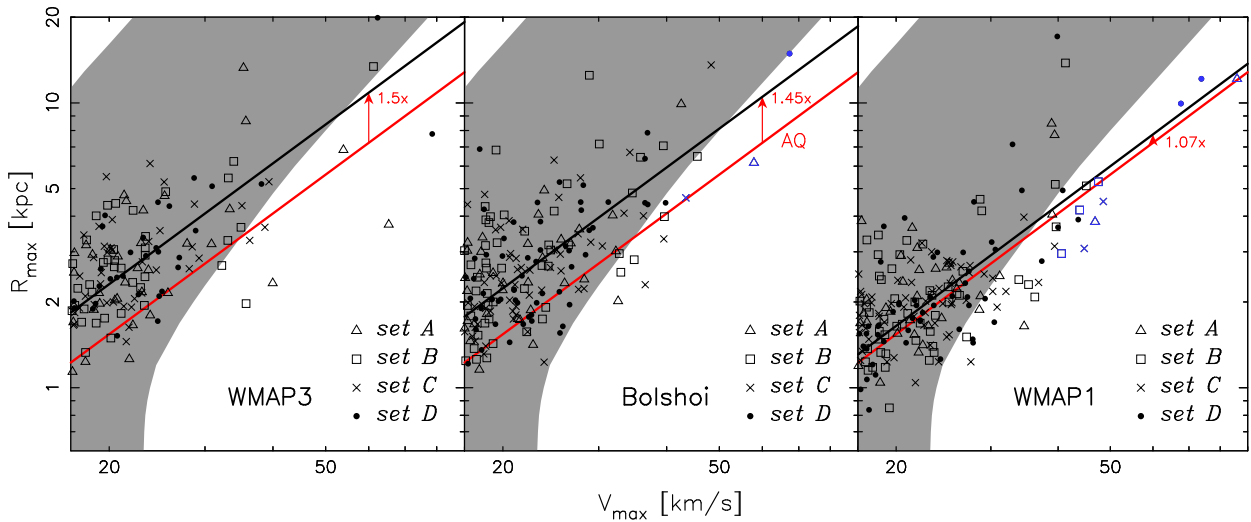}
\caption{Plots of $v_{max}$ and $R_{max}$ for subhaloes in the high 
  resolution CDM simulations
  for each set of cosmological parameters. 
  The shaded area shows the $2\sigma$ constraints for the
  bright Milky Way dwarfs from \citet{boy2011} assuming NFW
  profiles. The sloped red line shows the mean of the Aquarius
  subhaloes. Magellanic Cloud analogues in the Bolshoi and WMAP1 
  simulations are plotted in blue.\label{figVR}}
\end{figure*}

Figure~\ref{figVR} is a plot of $v_{max}$ and $R_{max}$ for subhaloes
in our high resolution CDM simulations.
\citet{boy2011} investigated what values of $v_{max}$ and $R_{max}$ of
NFW haloes \citep{nav1997} are consistent with the half-light
dynamical mass constraints of the bright MW dwarf spheroidals from
\citet{wolf2010}. Their $2\sigma$ confidence region is plotted as the
shaded regions in Figure~\ref{figVR}. 

We see there are many subhaloes that
lie in the range consistent with the MW dwarfs, but there are some
with $v_{max} > 20$~km~s$^{-1}$ that do not. 
These are the subhaloes
highlighted by \citet{boy2011} that are massive but have
central densities too high to host any of the MW dwarfs. 
However our WMAP3 and Bolshoi 
simulations have only 1-3 subhaloes per parent halo outside the shaded
zone of Milky Way satellites compared to 4-8 subhaloes for the 
WMAP1 simulations.
This is due to $R_{max}$ being shifted to higher values from WMAP1 
for the same values of $v_{max}$.

\citet{spr2008} show that the
logarithms of $v_{max}$ and $R_{max}$ for the Aquarius subhaloes have
a linear relationship. We estimate the equation of their fitting
line:
\begin{equation}\label{eq1}
\log R_{max} = 1.41 \log (v_{max}/14.72 \mbox{ km s$^{-1}$}),\\
\end{equation}
and plot this as the red line. 
We assumed a constant slope and performed least-squares fits 
to our subhaloes in each cosmology and plot these as the black lines.
The red arrowed lines show the shift in $R_{max}$ for each of our 
simulation sets 
compared to Aquarius.
Our simulations adopting WMAP1 parameters are in good agreement
with the Aquarius simulations, differing by only a factor of 1.07, 
but in Bolshoi and WMAP3 the subhaloes are offset to higher values 
of $R_{max}$ by factors of
1.45 and 1.50, respectively.

We compared the fit for each simulation set separately to the 
corresponding fit 
in the WMAP1 cosmology. We found the average scale in $R_{max}$ from WMAP1 
to Bolshoi is a factor of 1.35 and a factor of 1.40 for WMAP3, 
with a $1\sigma$ scatter of $\pm 0.10$ for each.

\begin{figure}
\includegraphics*[scale=0.34,angle=270]{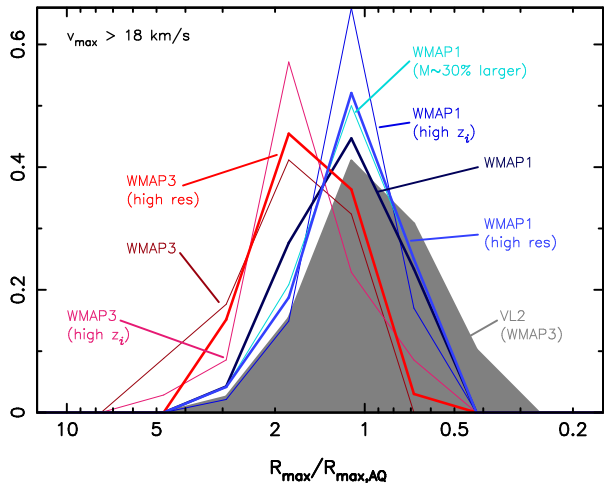}
\caption{Distribution functions of $R_{max}$ normalized to the
  Aquarius values for CDM subhaloes with $v_{max} > 18$~km~s$^{-1}$ in
  the WMAP1 and WMAP3 simulations of the \textit{set B} halo. 
  Simulations adopting WMAP3 parameters are plotted in red while WMAP1 
  simulations are plotted in blue.
  The offset between simulations is consistent with a cosmology 
  dependence and not
  on mass resolution, starting redshift, or mass of the host Milky Way halo.
  Solid gray area is the 
  distribution for Via Lactea-II subhaloes.\label{figCOMPARE}}
\end{figure}

To determine if factors other than the cosmology may be affecting 
the subhalo densities
we ran a series of tests on the \textit{set B} halo with the mass resolution
decreased a factor of 8 but the softening length kept the same as the
high resolution simulations. We ran a test adopting WMAP3 parameters
starting from the same initial redshift as the high resolution
simulation ($z_i=48$) and another test starting from a high redshift
($z_i=115$), comparable to the starting redshift of Aquarius
($z_i=127$). We also ran tests adopting the WMAP1 parameters. 
The Milky Way halo mass was about $30\%$ greater in this simulation so we
ran tests with the box size and mass resolution decreased
to give a halo mass similar to the WMAP3 tests. We ran small box tests
starting from the same low and high redshifts.

We examined applying velocity cuts of  
$v_{max} > 14-20$~km~s$^{-1}$ to the subhaloes. 
At smaller velocities the $R_{max}$ values for some subhaloes were 
inside the convergence radius 
satisfying the criterion of \citet{pow2003} 
and therefore affected by the resolution of the simulations.
In Figure~\ref{figCOMPARE} 
we normalize the values of $R_{max}$ for all subhaloes with 
$v_{max} > 18$~km~s$^{-1}$ 
to the Aquarius value of $R_{max}$ from Equation \ref{eq1}  
and present binned distributions for these subhaloes and
those of the Via Lactea-II (VL2) simulation ($z_i=104$).
We find consistent distributions between the low and high resolution
simulations showing our mass resolution and softening length are
sufficient to sample subhaloes with $v_{max} > 18$~km~s$^{-1}$. We also
find weak to no dependence on the starting redshift as the simulations 
started from $z_i = 115$ have distributions consistent with the corresponding
simulations started from $z_i = 48$. 
However, we do see a strong dependence on the 
cosmology as the WMAP3 simulations are offset to higher $R_{max}$ compared to 
WMAP1. The offset is only weakly dependent on the mass of the Milky
Way host as the WMAP1 simulations in the large and small boxes have 
nearly identical distributions.

We ran additional low resolution tests of the \textit{set B} halo 
adopting WMAP9, Bolshoi, and Planck1 parameters. 
These simulations also show offsets from WMAP1 but less than the 
WMAP3 tests (final column in Table~\ref{tabSETB}), 
as expected for the greater small scale power in these cosmologies. 
These tests show the subhalo
concentrations are largely determined by their formation time. 
As the small scale power increases formation occurs earlier and 
the subhaloes are more
concentrated at $z=0$. This is supported by examining the high
redshift data for these simulations. 
Table~\ref{tabSETB} gives the number of haloes with masses $>
2\times10^8 M_{\sun}$ and the average mass of the 12 largest haloes in 
the high resolution volume at $z=9$ in the test simulations of 
the \textit{set B} halo with mass resolution $7.34\times10^5 M_{\sun}$.
In the high resolution volume at
$z=9$ there are more than six times as many haloes with masses $>
2\times10^8 M_{\sun}$ in the WMAP1 simulation than in
WMAP3. Furthermore, the 12 most massive haloes are an average of four
times as massive in WMAP1 than WMAP3. This is evidence dwarf-scale
haloes are collapsing earlier and have more time to grow in a WMAP1
cosmology.

\begin{table}
\caption{Comparison of the low resolution CDM tests of the \textit{set B} halo with a common mass resolution. See text for an explanation of quantities in the columns.\label{tabSETB}}
\begin{center}
\begin{tabular}{l c c c}    
\hline
 Name & $N_{z=9}$ & $<M_{top 12}>$ & $\frac{R_{max}}{R_{max,WMAP1}}$ \\
 & $> 2\times10^8 M_{\sun}$ & [$10^9 M_{\sun}$] & \\
\hline
\hline
 WMAP1 & 378 & 2.939 & 1.0\\
 Planck1 & 239 & 1.982 & 1.06\\
 WMAP9 & 193 & 1.612 & 1.16\\
 Bolshoi & 149 & 1.375 & 1.20\\
 WMAP3 & 57 & 0.777 & 1.57\\
\hline
\end{tabular}
\end{center}
\end{table}

The distribution of VL2 subhaloes is also plotted in
Figure~\ref{figCOMPARE}. The VL2 simulation used WMAP3 cosmology but
its subhaloes have concentrations consistent with Aquarius. 
We hypothesize this is because the VL2 halo has a higher
redshift of formation than the mean for a WMAP3 cosmology.
Figure~\ref{figMH} shows 
our haloes generally have accreted less of their final mass at $a<0.5$ 
than the VL2 halo.
For example, at $a=0.25$ the VL2 halo has $23\%$ of its final
mass while our haloes have only $5-18\%$ of their final masses.
Further evidence comes from the halo concentration which is known to 
correlate with formation epoch.
We determined $M_{200}$ and $R_{200}$ 
($1.417\times10^{12} M_{\sun}$, 225.28 kpc) from the fit to the VL2 
density profile \citep{die2008} and calculate $c_{200}$ from 
Eqn~\ref{eqC200}. The concentration of VL2 is $10.7$, in contrast with 
the $6.6-8.7$ concentrations of our WMAP3 haloes.
VL2 is a $2.4 \sigma$ outlier in the WMAP3 simulations 
of \citet{mac2008} where
the average concentration of relaxed $10^{12} M_{\sun} h^{-1}$ haloes is 5.9.

\subsubsection{Velocity profiles}

A direct comparison of the subhalo circular velocity profiles 
to the half-light circular velocities of the observed dwarfs 
is desirable but is complicated by two effects. 
The circular velocity is a cumulative quantity and its profile is 
affected by the softening length to
greater distances than the density profile \citep{zol2012} making 
reliable inward 
extrapolation difficult.
Additionally, the hosts of the bright dwarfs are expected to be the 
largest subhaloes over the complete infall history of the subhalo 
population or the largest at the epoch of reionization. Many of 
these subhaloes will experience tidally stripped mass loss thereby 
reducing their $R_{max}$ sufficiently to become affected by the softening 
length. 
The largest subhaloes at present ($z=0$) are
generally subhaloes just beginning to infall as indicated by their 
large spatial extent \citep{and2013}. They are the least affected by 
stripping and therefore have the most reliable circular velocities.
Excluding Magellanic Cloud analogues from our simulations, 5-6 of 
the 10 subhaloes with greatest $v_{max}$ at $z=0$ are among the top 10 
with greatest $v_{infall}$ while 2-4 are among the top 10 with 
greatest $v_{max}$ at $z=9$.
Thus while we do not expect the largest subhaloes at $z=0$ to completely
match the observed dwarf population they are useful for 
illustrating the effects of cosmology on the too big to fail problem.

\begin{figure*}
\includegraphics*[scale=0.6,angle=270]{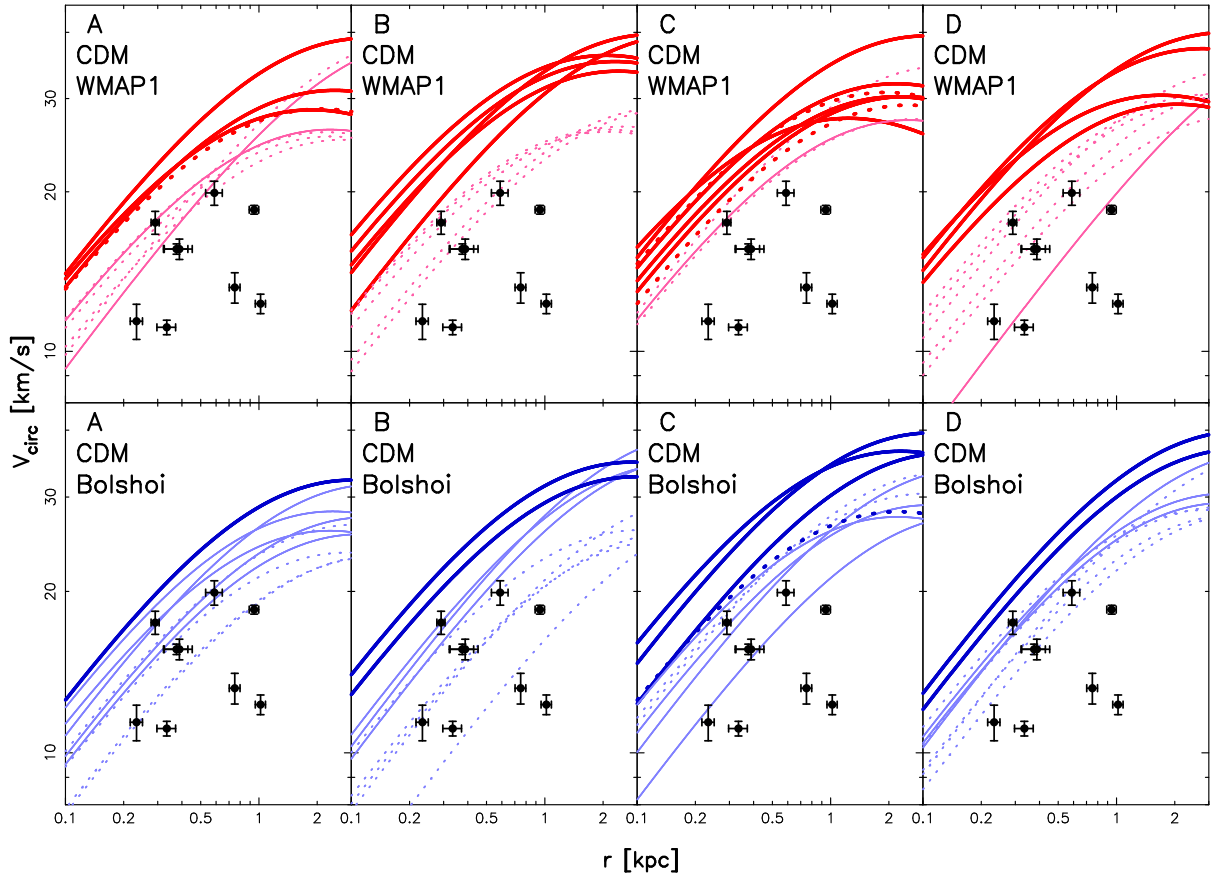}
\caption{NFW circular velocity profiles for the 10 subhaloes with largest
  $v_{max}$ at $z=0$ in each CDM simulation adopting WMAP1 cosmology 
  (\textit{top row}); and Bolshoi cosmology (\textit{bottom row}) after 
  filtering Magellanic Cloud analogues. Subhaloes denser than any observed 
  dwarf (points with error bars) are plotted in bold. Subhaloes that are 
  neither among the 10 with largest $v_{infall}$ or 10 largest $v_{max}$ at 
  $z=9$ are not expected to host a bright dwarf and are plotted with dotted 
  lines. Note that NFW profiles for the 10 subhaloes with largest $v_{max}$ 
  over their infall history select a few subhaloes with lower values of 
  $v_{max}$ and $R_{max}$ than shown here, further alleviating the discrepancy 
  with observations.
  \label{figVPRO}}
\end{figure*}

In Figure~\ref{figVPRO} we plot the NFW circular velocity profiles 
with $R_{max}$ and $v_{max}$ values of the 10 largest subhaloes in each 
CDM simulation adopting WMAP1 and Bolshoi cosmologies. 
The data points with error bars show the
circular velocities at half light radii of our bright Milky Way dwarfs
sample from \citet{wolf2010}.
While there is some halo-to-halo scatter the reduced densities and 
shift of the profiles to larger radii in the Bolshoi cosmology is 
dramatically clear.

\subsection{Warm Dark Matter}\label{sec:4}

Our results in the previous section show the discrepancy
between the largest subhaloes in CDM simulations and observations of
bright Milky Way dwarfs may largely be due to the adopted cosmological
parameters of the Aquarius simulation and that adopting parameters in
agreement with the most recent WMAP release would greatly alleviate this
problem. However we also saw that even a WMAP3 simulation like VL2 can
have massive satellites dynamically inconsistent with the bright
dwarfs implying a dependence on the formation history of the Milky Way
and its satellites. 
In this section we investigate the
effects warm dark matter has on the massive subhaloes.

\begin{figure*}
\includegraphics*[scale=0.64,angle=270]{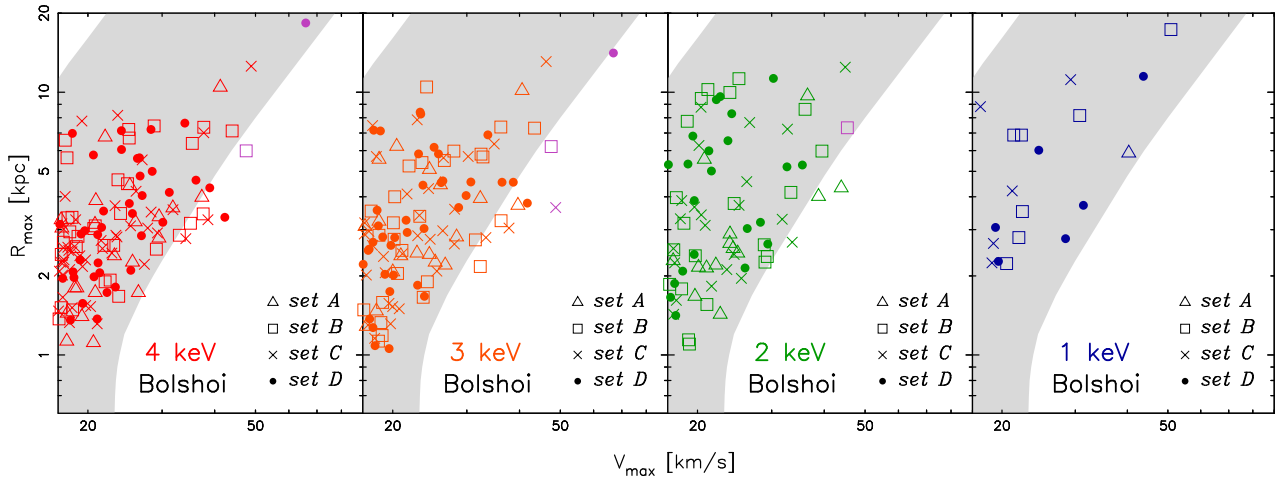}
\caption{Plots of $v_{max}$ and $R_{max}$ for subhaloes in the high 
  resolution WDM simulations adopting Bolshoi cosmological parameters. 
  The shaded area shows the $2\sigma$ constraints for the
  bright Milky Way dwarfs assuming NFW
  profiles. Magellanic Cloud analogues are colored purple.\label{fig4PS}}
\end{figure*}

Figure~\ref{fig4PS} is a plot of
$v_{max}$ and $R_{max}$ for subhaloes in each simulation set for 
each WDM cosmology.
Again we see there are many subhaloes that lie in the
area consistent with the MW dwarfs but there are some with 
$v_{max} > 20$~km~s$^{-1}$ that do not, however
the number of outliers decreases as the particle mass decreases. 
We find an average of 2~subhaloes per simulation are outside the allowed 
region decreasing to 1.5~per simulation in 3~keV, $<1$~in 2~keV, 
and 0~in 1~keV. 
We found an average of 2~subhalo outliers per Bolshoi CDM simulation 
demonstrating the minimal effect a 4~keV cosmology has on the densities.

The effects of WDM are a reduction in the total number of subhaloes as well as 
their circular velocities and an increase in their $R_{max}$.
We estimate the increase in $R_{max}$ by fitting equations of the form of 
Eqn~\ref{eq1}
to the WDM subhalo data and comparing to the fits for the corresponding 
CDM simulation. 
We find, for constant values of $v_{max}$, $R_{max}$ values are increased 
an average of $7\%$ in 4 keV, $15\%$ in 3 keV, $30\%$ in 2 keV, 
and $46\%$ in 1 keV; 
however, the small number of subhaloes in 1 keV makes it difficult to a
chieve a reliable estimate for this cosmology.

We estimate the effects of WDM on the circular velocities by
comparing the velocities at several radii in the range
$1-3$~kpc for subhaloes in WDM compared to the corresponding CDM simulation.
We find the subhaloes in 1~keV WDM have velocities up to
$60\%$ less than their CDM counterparts. This reduction decreases to
$20\%$ in 2~keV, $15\%$ in 3~keV, and only $10\%$ in 4~keV.

\subsubsection{Velocity profiles}

\begin{figure*}
\includegraphics*[scale=0.6,angle=270]{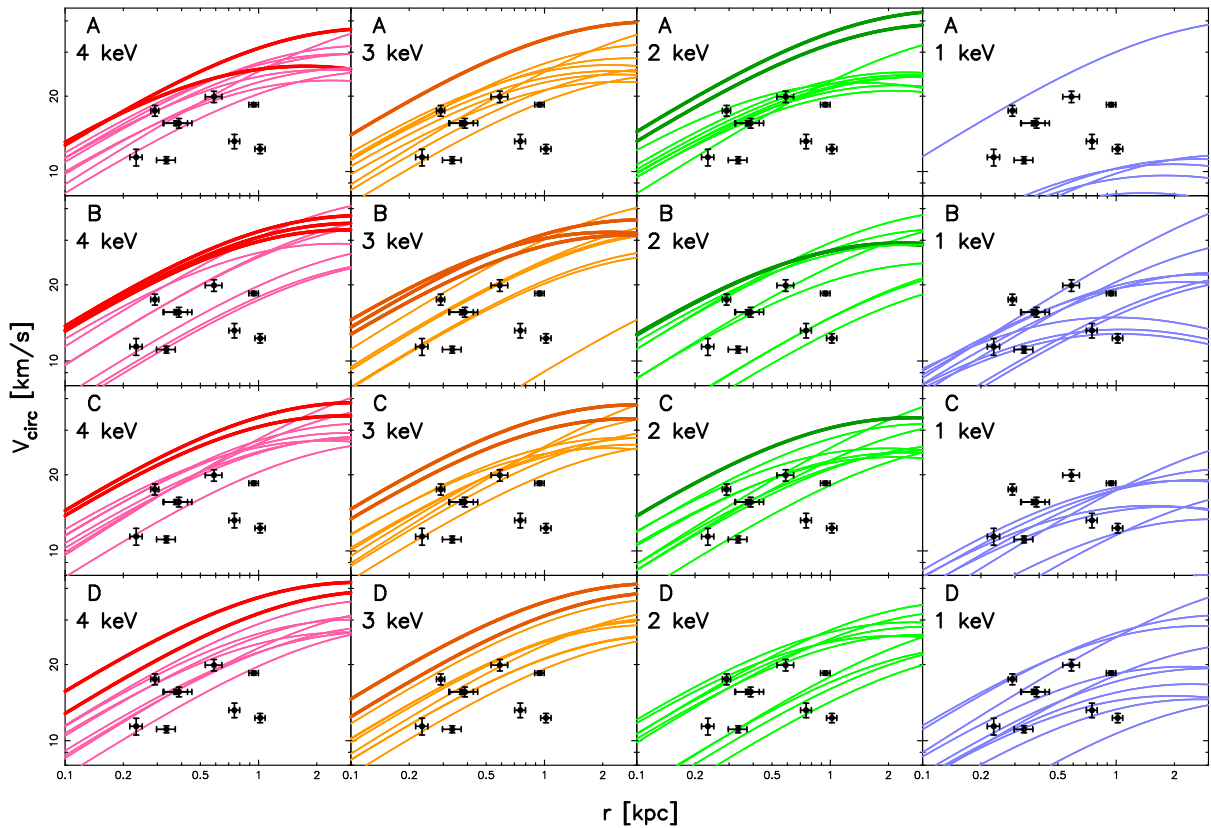}
\caption{NFW circular velocity profiles for the 10 subhaloes with largest
  $v_{max}$ at $z=0$ in each WDM simulation adopting Bolshoi cosmology. 
  Subhaloes denser than any observed dwarf (points with error bars) are 
  plotted in bold.
\label{fig7PS}}
\end{figure*}

Figure~\ref{fig7PS} shows the NFW circular 
velocity profiles of the 10
subhaloes with the largest $v_{max}$ at $z=0$ in our WDM simulations 
after excluding 
Magellanic Cloud analogues.

The subhalo profiles are severly affected in the 1 keV cosmology with
both the velocities and $R_{max}$ values showing large changes. 
The 1~keV simulations struggle to match the observations in number and 
density with only \textit{set D} managing to fit both.

Comparison to the CDM subhaloes plotted in Figure~\ref{figVPRO} shows 
some scatter among individual subhaloes. For example, a few subhaloes 
in \textit{set B} have increased density in WDM. In general, subhalo 
densities are significantly reduced in cosmologies warmer than 2~keV 
while at higher particle masses the effects are weak. This is in agreement 
with the single-halo simulations in \citet{sch2013} and \citet{lov2013a}.

\section{Discussion}

We found the concentrations and velocity profiles of subhaloes in CDM
simulations are dependent on the adopted cosmological parameters. We
tested and found little to no dependence on the starting redshift, 
the mass resolution, the mass of the parent halo, and the halo finding
software. 

A cosmological dependence is also seen in other published work of 
Milky Way-sized galaxies. 
The simulations of
\citet{sto2002} used similar parameters to Aquarius, ($\Omega_m$,
$\Omega_{\Lambda}$, $h$, $\sigma_8$, $n_s$) = (0.3, 0.7, 0.7, 0.9, 1),
and are well fit by Equation \ref{eq1}. \citet{dic2011} saw an
offset in their simulations using WMAP3 and WMAP5 parameters.
A dependence of substructure central densities on the cosmological parameters
is predicted in the work of \citet{zen2003} using the semianalytic 
model of \citet{bul2001a}.
The central densities are expected to reflect the mean density of the 
universe at the time of collapse. 
Adopting values for cosmological parameters that moves the formation
of small mass haloes to later epochs will result in less
concentrated subhaloes. 

Here we show how the subhalo densities can be
simply related to the power at their mass scale and therefore
dependent on both $\sigma_8$ and $n_s$.
The parameter $\sigma_8$ sets the power at a scale of 8~Mpc~$h^{-1}$
corresponding to a mass of about $2.5\times10^{14} M_{\sun}$. If the
mass of the largest satellites is about $10^{10} M_{\sun}$, the wave
number is $k_{sat} \sim 30 k_{8}$ where $k_8$ is the wave number
corresponding to 8~Mpc~$h^{-1}$. 
The change in $\sigma$ between WMAP3 and WMAP1 values of $n_s$ is given by:
\begin{equation}
\frac{k_{sat}}{k_{8}}^{(n_{s,WMAP3}-n_{s,WMAP1})/2} \sim 0.92.\\
\end{equation}
The change due to $\sigma_8$ is:
\begin{equation}
\frac{\sigma_{8,WMAP3}}{\sigma_{8,WMAP1}} \sim 0.82.\\
\end{equation}
The total change at the satellites scale is $0.92 \times 0.82 = 0.76$. 
This is also proportional to the change of the redshift of formation:
\begin{equation}
(1 + z_f)_{WMAP3} = 0.76 (1 + z_f)_{WMAP1}.\\
\end{equation}
The virial radius is proportional to $R_{max}$ at virialization and
the circular velocity at the virial radius is proportional to
$v_{max}$ at virilization and:
\begin{equation}
R_{vir} \propto v_{vir}(1 + z_f)^{-1.5}.\\
\end{equation}
Therefore we obtain the following scaling between cosmologies:
\begin{equation}
\frac{R_{max,WMAP3}}{R_{max,WMAP1}} = 0.76^{-1.5} = 1.51.\\
\end{equation}
Repeating this for the scaling between Bolshoi and WMAP1 cosmologies yields a
factor of 1.31.
From our simulations we derived average scaling factors of 1.40 and 1.35 
for WMAP3 and Bolshoi, respectively,
with a scatter of 0.10.
This is in good agreement with our rough calculation that assumes 
a mass of $10^{10}
M_{\sun}$ for the large satellites and neglects tidal effects that may
introduce a cosmology dependent change of the present values of
$R_{max}$ and $v_{max}$ from the values at virialization. We can
write an approximate general scaling ralation for $R_{max}$ at a fixed $v_{max}$:
\begin{equation}
R_{max} \propto (\sigma_8 5.5^{n_s})^{-1.5}.\\
\end{equation}
This equation gives a scaling of 1.24 between Planck1 and WMAP1.

We also investigated how the subhalo densities are affected in a range
of WDM cosmologies and quantified the reduction in circular velocity
at kpc scales. In previous work we have shown \citep{pol2011} that
the abundance of Milky Way satellites, including the ultra-faint
dwarfs discovered in the Sloane Digital Sky Survey, allow a 
lower limit of $2.3$~keV to be placed on the dark matter particle
mass. The work of \citet{lov2013a} favors a similar but slightly warmer 
limit of $1.6$~keV.
Lyman-$\alpha$ absorption by neutral hydrogen along the line of
sight to distant quasars over redshifts 2--6 probes the matter power
spectrum in the mildly nonlinear regime on scales 1--80 Mpc~$h^{-1}$.
Several authors have used Lyman-$\alpha$ data to provide independent
constraints on WDM with lower limits ranging from 1.7--4~keV
\citep{boy2009,vie2006,sel2006,vie2008,vie2013}.  Under these constraints we
expect the circular velocities of the largest satellites in WDM
to be affected by less than $20\%$, much less than the $60\%$ changes
seen in a 1~keV cosmology. We conclude that allowed WDM cosmologies
have only a mild effect on the density of massive Milky Way
satellites, that are instead most sensitive on the redshift of
formation of the Milky Way and the power at small scales given by
$\sigma_8$ and $n_s$.

While our simulations adopting Bolshoi cosmology reduced the number 
of ``too big to fail'' subhaloes in 3/4 of our Milky Way realizations 
from about four or five in WMAP1 to about one or two, none of our 
simulated Milky Ways are completely free of overdense subhaloes. 
Furthermore, the case of the VL2 halo demonstrates that large variation 
in average subhalo density is possible even in WMAP3 cosmologies.
\citet{pur2012} examined $10,000$ realizations of substructure for three 
host Milky Way masses from an analytic model. While their technique 
is only an approximation to direct simulation they find $\sim 10\%$ of 
their subhalo populations have no massive failures in a WMAP7 cosmology.
The Milky Way may thus simply be mildly atypical.
Interestingly, \citet{ham2007} show the Milky Way 
is deficient in stellar mass, disk angular momentum, and 
average iron abundance of stars in the Galactic halo at the $1\sigma$ level. 
Only $7\%\pm1\%$ of spiral galaxies with comparable rotation speeds 
have similar properties.
One way of explaining these discrepancies is to assume the Milky Way 
had a quiet accretion history without major merger events for the 
past $\sim 10$~Gyr. 
Figure~\ref{figMH} shows VL2 and our \textit{set B} and \textit{set C} 
haloes assemble $\sim 70\%$ of their mass by $z=1.5$ and may better 
represent the Milky Way than the other haloes, according to this model.
Opposite to expectations these haloes have the highest number of outliers.
However, \citet{pur2012} found selecting hosts for quiet accretion 
histories did not significantly increase the probability of consistency.

Our simulations assumed the dark matter was purely cold or purely
warm, but a mixture of the two is possible. The transfer function of
mixed dark matter is characterized by a step related to the particle
mass and a plateau at smaller scales related to the fraction of the
warm component. This could arise if the dark matter is composed of
multiple particle species or a single species containing warm and cold
primoridial momentum distribution components caused by separate
production stages, for example. \citet{boy2009} allowed for mixed
cold and warm dark matter in their analysis of Lyman-$\alpha$ forest
data. They find a particle mass of 1.1~keV is allowed if the WDM
fraction is less than 0.4 ($95\%$ confidence). 
Masses below 1 keV are allowed provided the
fraction of WDM is less than 0.35. 
\citet{and2013} examined a subhalo population in 
several mixed dark matter cosmologies. They show a range of models that
agree with Lyman-$\alpha$ constraints can be ruled out for failing to
produce subhaloes with sufficient density to match the observations, 
highlighting the usefulness and uniqueness of the Milky Way satellites 
as a probe of small-scale cosmology.

\section*{Acknowledgments}
We thank Owen Parry for help runnning {\sevensize SUBFIND}, 
J{\"u}rg Diemand for help accessing Via~Lactea~II data, and Michael 
Boylan-Kolchin for useful discussions. 
The simulations presented in this work were run on the 
{\sevensize DEEPTHOUGHT} computing cluster
at the University of Maryland College Park, the Cray XE6 {\sevensize GARNET} 
at the U.S. Army Engineer Research and Development Center, the retired Cray 
XE6 {\sevensize RAPTOR} and the SGI Ice X {\sevensize SPIRIT} at the 
U.S. Air Force Research Laboratory.
Basic research in astrophysics at NRL is funded by the
U.S. Office of Naval Research. 
EP acknowledges support under the Edison Memorial Graduate
Training Program at the Naval Research Laboratory. 
MR's research is supported by NASA
grant NNX10AH10G and NSF CMMI1125285.

\bibliographystyle{mn2e}
\bibliography{MMWS.v4}

\label{lastpage}
\end{document}